\documentclass[twocolumn,showpacs,preprintnumbers,amsmath,amssymb]{revtex4}

\usepackage{graphicx}
\usepackage{dcolumn}
\usepackage{bm}
\usepackage{amssymb}
\usepackage{epsfig}    
\usepackage{here}

\newcommand{\beq}{\begin{equation}}
\newcommand{\eeq}{\end{equation}}
\newcommand{\bq}{\begin{equation}}
\newcommand{\eq}{\end{equation}}
\newcommand{\ba}{\begin{array}}
\newcommand{\ea}{\end{array}}
\newcommand{\beqa}{\begin{eqnarray}}
\newcommand{\eeqa}{\end{eqnarray}}

\newcommand{\CP}{{\texttt C}{\texttt P}\,}

\newcommand{\nuBB}{$0\nu\beta\beta~$}

\newcommand{\hc}{ {\rm h.c.} }

\def\IM{\Im\mathfrak{m}}

\def\noi{\noindent}
\def\End{\end{document}}

\def\to{\rightarrow}
\def\To{\Rightarrow}

\def\dis{\displaystyle}
\def\f{\frac}
\def\ov{\overline}
\def\[{\left[}
\def\]{\right]}
\def\({\left(}
\def\){\right)}

\def\U1EM{U(1)_{\rm em}}

\def\leqq{\leqslant}
\def\geqq{\geqslant}

\def\ss{s_s^{~}}
\def\cs{c_s^{~}}
\def\sa{s_a^{~}}
\def\ca{c_a^{~}}
\def\sx{s_x^{~}}
\def\cx{c_x^{~}}
\def\sss{s_s^{2}}
\def\css{c_s^{2}}
\def\saa{s_a^{2}}
\def\caa{c_a^{2}}
\def\sxx{s_x^{2}}
\def\cxx{c_x^{2}}
\def\sh{\widehat{s}}
\def\TH{\theta}
\def\deg{\circ}

\def\[{\left[}
\def\]{\right]}
\def\dis{\displaystyle}

\def\N{{\cal N}}
\def\d{\delta}

\def\k{\kappa}
\def\D{\Delta}
\def\d{\delta}
\def\ep{\epsilon}
\def\N{{\cal N}}

\def\MM{\mathfrak{M}}
\def\m{\mathfrak{m}}  
\def\aa{\overline{a}}
\def\bb{\overline{b}}
\def\cc{\overline{c}}

\def\xaab{{\theta_{12}}}

\def\xabc{{\theta_{23}}}

\def\xaac{{\theta_{13}}}
\def\D{\Delta}
\def\d{\delta}

\def\ma{{m_1^{~}}}
\def\mb{{m_2^{~}}}
\def\mc{{m_3^{~}}}
\def\deg{\circ}

\def\s3{s_3^{~}}
\def\ep{\epsilon}
\def\epbar{\ov{\ep}_1^{~}}
\def\pp{\bar{p}}
\def\mD{$m_D^{~}$~}

\def\thisday{October, 2003}

\begin{document}

\preprint{\large hep-ph/0310278}

\title{%
Structure of Cosmological CP Violation via Neutrino Seesaw
}

\author{{\sc V. Barger}\,$^a$, \
        {\sc Duane A. Dicus}\,$^b$, \ 
        {\sc Hong-Jian He}\,$^b$,  \ 
        {\sc Tianjun Li}\,$^c$}
\affiliation{%
\vspace*{2mm}
$^{a}$\,Department of Physics, University of Wisconsin, Madison, WI~53706, USA 
\\ 
$^{b}$\,Center for Particle Physics and Department of Physics, 
        University of Texas at Austin, TX 78712, USA
\\
$^{c}$\,School of Natural Science, Institute for Advanced Study,
        Einstein Drive, Princeton, NJ~08540, USA
}%
\date{\thisday}

\begin{abstract}
\noindent
The cosmological matter-antimatter asymmetry 
can originate from \CP-violating
interactions of seesaw Majorana neutrinos via leptogenesis 
in the thermal phase of the early universe. 
Having the cosmological \CP~phase
for leptogenesis requires at least two right-handed
Majorana neutrinos. 
Using only the low energy neutrino observables,
we quantitatively reconstruct a minimal neutrino seesaw. 
We establish a general criterion for  
minimal seesaw schemes in which the cosmological \CP-phase
is {\it completely} reconstructed 
from the low energy \CP-phases measured by
neutrino oscillation and neutrinoless double-beta decay experiments.
We reveal and analyze two distinct classes of such  
minimal schemes that are shown to be highly predictive. 
Extension of our reconstruction formalism to
a three-heavy-neutrino seesaw is discussed.
\end{abstract}


\maketitle


\noindent  
{\bf 1. Introduction}
\vspace*{2.5mm}

From Big Bang Nucleosynthesis (BBN) the observed abundance of 
deuterium determines the baryon asymmetry in the universe to be 
$\eta_B^{\rm BBN} = (n_B^{~}-n_{\ov B}^{~})/n_{\gamma}^{~} 
                  = (6.1^{+0.7}_{-0.5})\times 10^{-10}$,\,
where $n_\gamma$ is the current photon number density\,\cite{BKLMS}. 
The precision of this asymmetry has been improved 
through measurements of the power spectrum of the 
cosmic microwave background radiation (CMB). 
From WMAP (Wilkinson Microwave Anisotropy Probe) data\,\cite{WMAP} 
the baryon density   \,$(\Omega_B^{~}h^2) = 0.0224\pm 0.0009$\,
is inferred, which corresponds to
\beq
\label{eq:CMB}
\eta_B^{\rm CMB}= \left(6.5^{+0.4}_{-0.3}\right)\times 10^{-10} \,.
\eeq

The standard model (SM) of particle physics 
fails to explain the observed baryon asymmetry because the 
Higgs mass bound from searches 
at the CERN Large Electron Positron (LEP) 
collider precludes a sufficiently strong first-order
electroweak phase transition\,\cite{SMEWPT}. 
Moreover, only a limited parameter range in the 
minimal supersymmetric extension of the SM (MSSM) 
is still viable for baryogenesis.

Alternatively, leptogenesis\,\cite{LG86} provides
a very attractive mechanism for realistic baryogenesis.
This possibility is especially well motivated by the discovery of
low energy neutrino oscillations in the past five years 
(cf. Refs.\,\cite{BMW-nuRev,other-nuRev} for recent reviews). 
The atmospheric neutrino data\,\cite{atm}
indicate $\nu_\mu\leftrightarrow\nu_\tau$ oscillations 
with nearly maximal mixing
\,$\TH_a \simeq 45^\deg$\, at the scale of
mass-squared-difference \,$\D_a\simeq 2\times 10^{-3}$\,eV$^2$,\,
while the solar and KamLAND data\,\cite{sun} single out 
$\nu_e\leftrightarrow \nu_\mu$ oscillations with
large mixing angle (LMA) solution of \,$\TH_s\simeq 32^\deg$\, 
and the scale of mass-squared-difference 
\,$\D_s\simeq 7\times 10^{-5}$\,eV$^2$.\,
The CHOOZ and Palo Verde
long baseline reactor experiments (in combination
with the mass range of the atmospheric oscillation data)
constrain the $\nu_e\to\nu_\tau$ transition to be fairly small,
\,$\TH_x \lesssim 13^\deg$ ($\sin\TH_x\lesssim 0.22$)\,
at 95\%\,C.L.\,\cite{chooz}, which bounds Dirac \CP-violation. 
The seesaw mechanism\,\cite{seesaw} 
provides a most natural explanation for
the tiny masses of active neutrinos, and the
predicted Majorana nature induces two independent Majorana
phases at low energy, relevant to  
neutrinoless double-beta ($0\nu\beta\beta$)
decay experiments\,\cite{0nuBB}.
In addition,  the seesaw mechanism further allows the natural realization
of successful leptogenesis, with the light neutrino mass range
compatible with the current low energy oscillation 
data\,\cite{Buch}.

A crucial ingredient for leptogenesis is the generation of lepton asymmetry via
\CP-violations in the neutrino seesaw mechanism\,\cite{LG86}.
Given the excitement from the various current 
and upcoming neutrino experiments 
(especially Superbeams and next generation \nuBB decay experiments),
an intriguing possibility is that 
{\it the baryon asymmetry of our universe may be
connected to the low energy neutrino \CP-violation 
phenomenon and thus highly testable.}
We note that a seesaw sector with $k$ right-handed neutrinos will 
contain \,$3(k-1)$\, independent \CP-phases ($k=1,2,3$). 
The required \CP-violation for leptogenesis can naturally arise from the
\CP-phases in the seesaw sector, provided $k\geqq 2$.
The Minimal Neutrino Seesaw Models (MNSMs) for leptogenesis
contain two heavy Majorana neutrinos $(N_1,\,N_2)$.
Although a link between high energy leptogenesis 
and low energy neutrino \CP-violations
is not guaranteed in general\,\cite{Reb}, 
in this Letter we will quantitatively 
derive a {\it minimal criterion} for the MNSMs under which
the high energy leptogenesis \CP-asymmetry can be {\it completely} 
reconstructed from the low energy neutrino \CP-violations.
We reveal and analyze two general classes of such MNSMs, 
and derive the predictions
for the allowed ranges of low energy neutrino $\CP$-observables  
that can generate the observed baryon asymmetry. 
Finally the extension of our reconstruction formalism to analyzing
the leptogenesis in the three neutrino 
$(N_1,\,N_2,\,N_3)$ seesaw is discussed.

\vspace*{3mm}
\noindent 
{\bf 2. Leptogenesis CP Asymmetry in the Minimal\\
 Neutrino Seesaw}
\vspace*{1.5mm}

A lepton asymmetry can be dynamically generated
in an expanding universe, provided that all three of 
Sakharov's conditions\,\cite{Sak} are realized: 
(i) the heavy Majorana neutrinos $N_j$ decay into a lepton-Higgs pair
$\ell H$ and its \CP-conjugate pair $\ov{\ell} H^*$
with different partial widths, thereby violating lepton number; 
(ii) \CP-violation arises from phases in the seesaw sector;
(iii) cosmological expansion causes departure from the thermal equilibrium.
As the universe cools down, the $N_j$'s drop out of equilibrium and their
decays generate a \CP asymmetry,
\beq
\ep_j =\dis\f{~\Gamma[N_j\to\ell H]-\Gamma[N_j\to\ov{\ell}H^*]~}
             {~\Gamma[N_j\to\ell H]+\Gamma[N_j\to\ov{\ell}H^*]~}.
\eeq
This leads to a lepton asymmetry, 
\beq
\label{eq:YL}
\dis Y_L = \f{n_L^{~}-n_{\ov{L}}}{s}
         = \sum_j \f{\ep_j^{~}\kappa_j^{~}}{g_{*j}^{~}} \,,
\eeq
where $g_{*j}^{~}$ represents the relativistic degrees of 
freedom which contribute to the entropy $s$ at a
temperature of the order of the lightest heavy Majorana neutrino 
mass $M_j$; in the SM and MSSM, for instance,    
$g_{*1}^{~} = 106.75$ and $g_{*1}^{~} = 228.75$, 
respectively\,\cite{KT}.  
[Here and below we use SM\,(MSSM) to denote
 the usual SM\,(MSSM) with
 an addition of the neutrino seesaw sector.]
In (\ref{eq:YL}),
$\kappa_j^{~}$ describes the wash-out of the asymmetry $\ep_j^{~}$ 
due to the various lepton-number-violating processes.
The wash-out factor $\kappa_j^{~}$ can be computed from the 
full Boltzmann equations.
Analytically $\kappa_1^{~}$ can be approximated as\,\cite{KT}
\beq
\dis
\kappa_1^{~} 
\,\simeq\, 0.3\[\f{10^{-3}\,{\rm eV}}{\widetilde{m}_1^{~}}\]
              \[\ln\f{\widetilde{m}_1^{~}}{10^{-3}\,{\rm eV}}\]^{-3/5} ,
\eeq
for $\,\widetilde{m}_1^{~} = (10^{-2}\mbox{--}10^3)$\,eV,
where \,$\widetilde{m}_1^{~} = (m_D^\dag m_D^{~})_{11}/M_1$\,
[cf. (\ref{eq:Lseesaw})-(\ref{eq:mD}) below for definitions 
of $m_D^{~}$ and $M_1$)].
More accurate empirical formulas for $\k_1$ fitted to 
the exact solution were also derived\,\cite{king}.  
The nonperturbative sphaleron interactions 
violate $B+L$ but preserve $B-L$ \cite{tHooft},
so that the lepton asymmetry $Y_L$ is partially transmitted to 
a baryon asymmetry $Y_B$\,\cite{Turner}
\beq
\label{eq:YB}
\dis
Y_B   
\,=\,\f{n_B^{~}-n_{\ov B}}{s}
\,=\, \f{\xi}{\xi-1}\,Y_L 
\,=\, \f{\xi}{(\xi-1)} \sum_j \f{\,\ep_j^{~}\kappa_j\,}{g_{*j}^{~}} 
\,,
\eeq
where \,$Y_B\simeq \eta_B^{~}/7.04$\,.\, The parameter 
$\xi$ depends on the other processes in equilibrium; it 
is expressed as \,$\xi = (8N_F^{~}+4N_H^{~})/(22N_F^{~}+13N_H^{~})$\,
with $N_F^{~}$ the number of fermion generations and $N_H^{~}$ the
number of Higgs doublets. 
Thus, given the particle contents of the SM (MSSM), 
we have \,$\xi =28/79\,(8/23)$\,.

\vspace*{1.5mm}
The minimal seesaw Lagrangian can be written as
\beq
\label{eq:Lseesaw}
\ba{l}
\\[-6mm]
\dis
{\cal L}_{ss} = -\ov{\ell_L}M_\ell \ell_R
                  -\ov{\nu_L^{~}}m_D^{~} \N   
                  +\f{1}{2}\,\ov{\N^c}^T\!M_R \,\N +\hc ,
\ea
\eeq
where the light neutrinos   
$\nu_L^{~}=(\nu_e,\,\nu_\mu,\,\nu_\tau)^T$  are in the flavor eigenbasis, 
while the charged leptons $\ell=(e,\,\mu,\,\tau)^T$ and 
heavy Majorana neutrinos $\N = (N_1,\,N_2)^T$ 
are in their mass eigenbasis so the mass matrices
$M_\ell ={\rm diag}(m_e,\,m_\mu,\,m_\tau)$ and 
$M_R ={\rm diag}(M_1,\,M_2)$ are diagonal and real. 
The $3\times2$ Dirac
mass matrix $m_D^{~}$ contains all 3 independent \CP-phases in 
the seesaw sector, and can be generally written as
\beq
\label{eq:mD}
m_D^{~} ~= \dis 
\left\lgroup 
\ba{cc} 
a~ & a' \\[2mm]
b~ & b' \\[1mm]
c~ & c' 
\ea 
\right\rgroup
=  
\left\lgroup 
\ba{cc} 
\zeta_1\aa~ & \zeta_2\,\aa' \\[1.4mm]
\zeta_1\bb~ & \zeta_2\,\bb' \\[1mm]
\zeta_1\cc~ & \zeta_2\,\cc' 
\ea 
\right\rgroup \! ,  
\eeq
where 
\,$\zeta_1 \equiv \sqrt{m_0^{~}M_1}$\,  
and 
\,$\zeta_2 \equiv \sqrt{m_0^{~}M_2}$\,
with 
\,$m_0^{~}\equiv\sqrt{\D_a}$.  
In (\ref{eq:mD}) we have defined, for convenience, the 6 dimensionless parameters,
$\,(\aa,\,\bb,\,\cc )    \equiv (a,\,b,\,c)/\sqrt{m_0^{~}M_1}$   and
$\,(\aa',\,\bb',\,\cc' ) \equiv (a',\,b',\,c')/\sqrt{m_0^{~}M_2}$,\,
which contain all 3 independent \CP-phases.
Integrating out the heavy neutrinos from the seesaw Lagrangian
(\ref{eq:Lseesaw}) we arrive at the symmetric $3\times3$ Majorana mass 
matrix for active neutrinos
\beq
\label{eq:Mnu}
\ba{lll}
\MM_\nu & \simeq & m_D^{~}M_R^{-1} m_D^T
\\[3mm] 
        &   =    & 
m_0^{~} 
\left\lgroup
\ba{ccc}
\aa^2+\aa'^2 & \aa\bb+\aa'\bb' & \aa\cc+\aa'\cc' 
\\[2mm]
             & \bb^2+\bb'^2    & \bb\cc+\bb'\cc'
\\[2mm]
             &                 & \cc^2+\cc'^2 
\ea
\right\rgroup \!.
\ea
\eeq
Note that $\MM_\nu$ can be completely expressed in terms
of 6 dimensionless parameters 
$(\aa,\,\bb,\,\cc )$ and $(\aa',\,\bb',\,\cc' )$,
up to an overall mass scale parameter $m_0^{~} = \sqrt{\D_a}$\,.

We can readily verify that  
$\det\(\MM_\nu\) = 0$, so one of the
three neutrino mass eigenvalues $(m_1,\,m_2,\,m_3)$ must be zero. 
Therefore the minimal
seesaw predicts a {\it hierarchical neutrino mass spectrum} and 
excludes a nearly degenerate mass spectrum.
The hierarchical mass spectra include the 
Normal Hierarchy (NH) and Inverted Hierarchy (IH),
\beq
\label{eq:NH-IH}
\ba{rl}
\hspace*{-14mm}
{\rm NH\!:}~~~~ &  0=m_1 \,<\, m_2 \,\ll\, m_3 \,,
\\[3mm]
\hspace*{-14mm}
{\rm IH\!:}~~~~ &  m_1 \,\gtrsim\, m_2 \,\gg\, m_3=0 \,,
\ea
\eeq 
where for NH we have
\,$ {m_2} = m_0^{~}\sqrt{r},~ 
    {m_3} = m_0^{~}\sqrt{1+r}$;
and for IH we have
${m_1} = m_0^{~} \sqrt{1+r},~ 
 {m_2} = m_0^{~}$.\,
Here $r$ is defined as the ratio  $r\equiv \D_s /\D_a$,
which is constrained by the oscillation data to the range
$1.9\times 10^{-2} \lesssim r \lesssim 7.4\times 10^{-2} $ (95\%\,C.L.), 
with a central value $r\simeq 0.036$.

Now considering thermal leptogenesis with $M_1 \ll M_2$ for simplicity, 
only the decays of the lightest right-handed neutrino 
$N_1$ are relevant for generating the final baryon asymmetry $Y_B$
in (\ref{eq:YB})  because the asymmetry caused by
$N_2$ decays is washed out by the lepton-number-violating processes
involving $N_1$'s that are more abundant at the high temperature
$T \sim M_2 \gg M_1$. But when temperature $T\sim M_1\ll M_2$ is reached,
 the heavier $N_2$'s 
have already decayed and the asymmetry from
$N_1$ decays is preserved.  
Thus, under this mechanism, 
only the term $\ep_1\kappa_1/g_{*1}$ in (\ref{eq:YB}) 
will contribute. 
We explicitly derive the \CP-asymmetry parameter 
$\ep_1^{~}$ to be
\beq
\label{eq:Epsilon1}
\ba{l}
\ep_1^{~} = \dis\f{\,m_0^{~}M_2\,}{8\pi}\(\f{\sqrt{2}}{v\sin\!\beta}\)^2\!
               F\!\(\f{M_2}{M_1}\)\,\ov{\ep}_1^{~} \,,
\\[5mm]
\ov{\ep}_1^{~}  \equiv\!  \dis\f{\Im\m\[\((m_D^\dag m_D^{~})_{12}\)^2\]}
                            {(m_0^{~}M_2)(m_D^\dag m_D^{~})_{11}} 
=\!\dis\f{\Im\m\[(\aa^*\aa'+\bb^*\bb'+\cc^*\cc')^2\]}
            {|\aa|^2+|\bb|^2+|\cc|^2} ,
\ea
\eeq
where \,$\sin\!\beta = 1\,(< 1)$\, for SM (MSSM), and
the function $F(x)$ is given by\,\cite{Buch}
\beq
\label{eq:f}
F(x) = \dis\left\{ 
\ba{lr}
\dis x\[1-(1+x^2)\ln\f{1+x^2}{x^2} +\f{1}{1-x^2}\], & ({\rm SM}),
\\[4mm]
\dis x\[-\ln\f{1+x^2}{x^2} +\f{2}{1-x^2}\], & ({\rm MSSM}).
\ea \right.
\eeq
The \CP-asymmetry parameter $\ov{\ep}_1$
in (\ref{eq:Epsilon1}) is solely expressed as a function
of the six dimensionless parameters 
$(\aa,\,\aa',\,\bb,\,\bb',\,\cc,\,\cc')$
introduced for the Dirac mass matrix $m_D^{~}$ in Eq.\,(\ref{eq:mD}).
For the case $M_1\gg M_2$, the relevant \CP-asymmetry is \,$\ep_2^{~}$
from $N_2$ decays and can be independently analyzed in the same way
as the reconstruction analysis of $\ep_1^{~}$ below.

\vspace*{5mm}
\noindent 
{\bf 3. Structure of Cosmological CP Violation 
        from Low Energy Reconstruction
}
\vspace*{2.5mm}

\noindent 
\underline{{\Large $\bullet$}\,{\it Reconstructing 
the $3\times 3$ Neutrino Mass Matrix~}}
\vspace*{3mm}

Consider a generic $3\times3$ symmetric Majorana mass matrix
$\MM_\nu$ for 3 light flavor-neutrinos $(\nu_e,\,\nu_\mu,\,\nu_\tau)$,
\beq
\label{eq:Mnulow}
\MM_\nu = \!
\left\lgroup 
\ba{ccc}
m_{ee}     &  m_{e\mu}    &  m_{e\tau}   \\[2mm]
           &  m_{\mu\mu}  &  m_{\mu\tau} \\[2mm]
           &              &  m_{\tau\tau}
\ea
\right\rgroup 
\equiv  \,
m_0^{~}\!\!
\left\lgroup 
\ba{ccc}
 A & C & D \\[1mm]
   & B & E \\[1mm]
   &   & F
\ea 
\right\rgroup \!.
\eeq
The mass matrix (\ref{eq:Mnulow}) contains 
nine independent real parameters, which can be
equivalently chosen as three mass eigenvalues 
$(\ma,\,\mb,\,\mc )\geqq 0$,
three mixing angles
$(\xaab ,\,\xabc ,\,\xaac )$,
and three CP-violation phases 
$(\d ; \phi,\,\phi')$ with $\d$ the usual Dirac phase
and $(\phi,\,\phi')$ the Majorana phases (irrelevant to
neutrino oscillations).
For MNSMs we have $\det (\MM_\nu) =0$ 
so that there are only two nonzero mass-eigenvalues [cf.\ (\ref{eq:NH-IH})].
The neutrino mixing matrix\,\cite{PMNS} is 
$V\equiv UU'$,  which diagonalizes  $\MM_\nu$
via $V^T\MM_\nu V = \MM_\nu^{\rm diag}$, and
contains six parameters (three rotation angles and three phases). 
The unitary matrix $V$ can
be decomposed into a matrix $U$ (\`{a} la Cabibbo-Kobayashi-Maskawa) and 
a diagonal matrix $U'$ with two independent Majorana phases, i.e.,
\beq
\label{eq:U}
U=
\left\lgroup
\ba{ccc}
\cs\cx    &   -\ss\cx         &   -\sx e^{i\d} \\[2mm]
\ss\ca - \cs\sa\sx e^{-i\d}   &  \cs\ca +\ss\sa\sx e^{-i\d} 
& -\sa\cx  \\[2mm]
\ss\sa + \cs\ca\sx e^{-i\d}   &  \cs\sa - \ss\ca\sx e^{-i\d} 
& \ca\cx
\ea
\right\rgroup \,.
\eeq
where for convenience we have introduced the notation
$\,
(\TH_s,\,\TH_a,\,\TH_x) \equiv
(\xaab ,\,\xabc ,\,\xaac )
$,\,
with 
\,$(s_\alpha^{~},\,c_\alpha^{~})\equiv (\sin\TH_\alpha,\,\cos\TH_\alpha)$\,
for  $\alpha=s,a,x$.\,   
The matrix $U'$ can be parametrized as,
    \,$U' = {\rm diag}(e^{-i\phi'/2},\, e^{-i\phi/2},\, 1)$\, for NH schemes
and \,$U' = {\rm diag}(1,\, e^{-i\phi/2},\, e^{-i\phi'/2})$\, for IH schemes.
[Our convention of $V$ is related to that of particle data book
via \,$(\TH_{s,a,x},\,\d,\,\phi,\,\phi' )\to 
      -(\TH_{s,a,x},\,\d,\,\phi,\,\phi' )$\,.]
Because of the mass spectrum of the minimal seesaw in Eq.\,(\ref{eq:NH-IH}), 
we see that in the above parametrization of $U'$ only a single
Majorana phase $\phi$ is physically relevant.
From the mass diagonalization, 
we can reconstruct the neutrino
mass matrix $\MM_\nu$ via the relation
\beq
\label{eq:MnuR}
\MM_\nu  =  V^\ast \,\MM_\nu^{\rm diag}\, V^\dagger \,,
\eeq
which gives
\beq
\label{eq:Mnuij}
\ba{lcl}
m_{ee} &\!\!=\!\!&
\cxx
\[
\css m_1'+\sss m_2'
\] + \pp^2\sxx m_3' \,,
\\[1.6mm]
m_{\mu\mu}  &\!\!=\!\!&
(\ss\ca - {p}\,\cs\sa\sx )^2m_1' +
(\cs\ca + {p}\,\ss\sa\sx )^2m_2' 
\\[1.6mm]
&& 
+\, \saa\cxx m_3' \,,
\\[1.6mm]
m_{\tau\tau}  &\!\!=\!\!&
(\ss\sa + {p}\,\cs\ca\sx )^2m_1'  +
(\cs\sa - {p}\,\ss\ca\sx )^2m_2' 
\\[1.6mm]
&& 
+ \, \caa\cxx m_3' \,,
\\[1.6mm]
m_{e\mu}  &\!\!=\!\!&
\cx \[
\ss\cs\ca (m_1' - m_2') - 
{p}\,\sa\sx (\css m_1' + \sss m_2')   \right. 
\\[1.6mm]
&& \left.
+ \,\pp\, \sa\sx m_3'
\] \,,
\\[1.6mm]
m_{e\tau}   &\!\!=\!\!&
\cx \[
\ss\cs\sa (m_1' - m_2')
+ {p}\,\ca\sx (\css m_1' + \sss m_2') \right.
\\[1.6mm]
&&  \left.
- \,\pp\, \ca\sx m_3'
\],
\\[2mm]
m_{\mu\tau}   &\!\!=\!\!&
(\ss\sa \!+\! {p}\,\cs\ca\sx )
(\ss\ca \!-\! {p}\,\cs\sa\sx )m_1' 
\\[1.6mm]
&& 
\hspace*{-2mm}
+(\cs\sa \!\!-\! {p}\,\ss\ca\sx )(\cs\ca \!\!+\! {p}\,\ss\sa\sx )m_2' 
          \!\!-\! \sa\ca\cxx m_3' .
\ea
\eeq
For NH schemes 
$(m_1',\,m_2',\,m_3') \equiv 
  (0,\,m_2^{~}q,\,m_3^{~} ) = m_0^{~}(0,\,\sqrt{r}q,\,\sqrt{1+r} ) 
 $\,,\,
for IH schemes
 $(m_1',\,m_2',\,m_3') \equiv 
  (m_1^{~},\,m_2^{~}q,\,0 ) = m_0^{~}(\sqrt{1+r},\, q,\, 0 ) 
 $\,,\,
and the two relevant \CP-phase factors are
$(p,\,q) = ({\bar p}^\ast,\,q)  
          = (e^{i\d},\,e^{i\phi})$.

Unlike most other bottom-up approaches, 
our reconstruction formalism solely relies on the low energy neutrino
observables without any extra ansatz about the specific forms of the
neutrino mass matrix (\ref{eq:Mnulow}), and is thus completely
general for a given seesaw sector.

\vspace*{5mm}
\noindent
\underline{{\Large $\bullet$}\,{\it Reconstruction Theorem 
in Minimal Neutrino Seesaw}}
\vspace*{3mm}

From (\ref{eq:Mnu}) and (\ref{eq:Mnulow}), we derive the 
reconstruction equation:
\beq
\label{eq:RCE}
\left\lgroup
\ba{ccc}
\aa^2+\aa'^2 & \aa\bb+\aa'\bb' & \aa\cc+\aa'\cc' 
\\[2mm]
             & \bb^2+\bb'^2    & \bb\cc+\bb'\cc'
\\[2mm]
             &                 & \cc^2+\cc'^2 
\ea
\right\rgroup
=
\left\lgroup 
\ba{ccc}
 A & C & D \\[1mm]
   & B & E \\[1mm]
   &   & F
\ea 
\right\rgroup \!.
\eeq
This equation contains 6 (complex) conditions
among which 5 are independent and the sixth condition is redundant because
of $\det(\MM_\nu)=0$ in the minimal seesaw. Accordingly, the
property $\det(\MM_\nu)=0$ also makes one of the 6 
elements $(A,B,C,D,E,F)$ to be dependent. 
Therefore we can analytically solve (reconstruct) 5 out of the 6 dimensionless 
seesaw parameters 
$(\aa,\,\aa',\,\bb,\,\bb',\,\cc,\,\cc')$ of $m_D^{~}$
in terms of the low energy neutrino observables contained in $(A,B,C,D,E,F)$.
For $A\neq 0$, we can exactly resolve (\ref{eq:RCE}) 
in terms of $\aa$ or $\aa'$,
\beq
\label{eq:RCS1}
\ba{ll}
\aa' & = \dis \sh_{a'}\sqrt{A-\aa^2\,} \,,~~~{\rm or},~~~  
       \aa = \dis \sh_{a}\sqrt{A-\aa^{\prime 2}\,}\,, \
\\[2mm]
\bb  & = \dis \f{1}{A}\[\aa\, C - \sh_{b'}\aa'\sqrt{AB-C^2}\] ,
\\[3mm]
\bb' & = \dis \f{1}{A}\[\aa'\,C + \sh_{b'}\aa \sqrt{AB-C^2}\] ,
\\[3mm]
\cc  & = \dis \f{1}{A}\[\aa\, D - \sh_{c'}\aa'\sqrt{AF-D^2}\] ,
\\[3mm]
\cc' & = \dis \f{1}{A}\[\aa'\,D + \sh_{c'}\aa \sqrt{AF-D^2}\] ,
\\[3mm]
E    & = \dis \f{1}{A}\[CD + \sh_{b'}\sh_{c'}\sqrt{(AB-C^2)(AF-D^2)\,}\] ,
\ea
\eeq
where
$(\sh_a,\,\sh_{a'},\,\sh_{b'},\,\sh_{c'})=\pm 1$.
We see that with a single input of either $\aa$ or $\aa'$ the other
5 dimensionless parameters in $m_D^{~}$ are completely determined
from the low energy neutrino observables contained
in $(A,B,C,D,F)$; the last equation is a consistency condition
due to $\det(\MM_\nu)=0$ that will also fix the sign convention
of $\sh_{b'}\sh_{c'}$ from the reconstructed $3\times3$
low energy neutrino mass matrix $\MM_\nu$ in (\ref{eq:Mnulow}) and
(\ref{eq:Mnuij}). 
Noting that the reconstruction equation (\ref{eq:RCE}) is invariant
under the joint transformations 
%
$\,
\aa  \leftrightarrow \bb  ,~
\aa' \leftrightarrow \bb' ,~
A    \leftrightarrow B    ,~
D    \leftrightarrow E ,\,
$ 
%
we can readily resolve (\ref{eq:RCE}) in terms of $\bb$ or $\bb'$,
for $B\neq 0$. 
Similarly, using the invariance of (\ref{eq:RCE}) under the exchanges
%
$\,
\aa  \leftrightarrow \cc  ,~
\aa' \leftrightarrow \cc' ,~
A    \leftrightarrow F    ,~
C    \leftrightarrow E ,\, $
%
we can reexpress the solutions in terms of  $\cc$ or $\cc'$ instead,
for $F\neq 0$, and so on.
The exact solution  (\ref{eq:RCS1})  
explicitly shows that from the low energy neutrino observables
in $\MM_\nu$ we can reconstruct 5 out of the 6 dimensionless
parameters in (\ref{eq:mD}) for $m_D^{~}$.  In the minimal seesaw
$m_D^{~}$ contains a maximal 6 \CP-phases among which only 3 
combinations are independent, and one independent
combination gives the required \CP-phase for leptogenesis.
Hence, we can formulate a Reconstruction Theorem for leptogenesis:

{\it In the minimal neutrino seesaw models (MNSMs) with the
leptons and heavy Majorana neutrinos in their 
mass-eigenbasis, five out of six dimensionless parameters 
$(\aa,\,\aa',\,\bb,\,\bb'\!,\,\cc,\,\cc')$ in $m_D^{~}$  
can be completely reconstructed from
the low energy neutrino observables. A complete reconstruction
can be realized in a class of MNSMs where one of the six entries in
$m_D^{~}$ is not independent. The simplest such MNSMs can be classified
into Type-I and Type-II, where Type-I has 1-texture-zero
and Type-II has 1-equality between two entries of
$m_D^{~}$. Consequently, for Type-I and -II, 
all \CP-phases in $m_D^{~}$ are fully
reconstructable from low energy, so that a unique
link is realized
between the high energy leptogenesis \CP-asymmetry
and the low energy neutrino \CP-violation observables.}

In fact the requirement of 1-texture-zero or 1-equality is the
{\it Minimal Criterion} for a complete reconstruction of the seesaw sector
(up to two mass parameters $M_{1,2}$ of $N_{1}$ and $N_2$). 
We note that in the above reconstruction
the \CP-phase distribution in all 6 entries of \mD is also fully fixed,
and there is no more freedom for rephasing of lepton-doublets in the
Yukawa interaction or Dirac mass term because we have made use of the
rephasing of all three lepton-doublets to derive the standard 
parametrization of the MNSP mixing matrix $V=UU'$ \cite{PMNS}  in
the leptonic charged currents such that 
only 3 independent \CP-phases $(\d,\,\phi,\,\phi')$ enter our low energy
formulation (where one of the two Majorana phases $(\phi,\,\phi')$
becomes irrelevant in the MNSMs). 
A unique feature of our reconstruction formalism is that
it does not introduce any extra unphysical \CP-phase beyond the
standard MNSP mixing matrix $V$ \cite{PMNS}, so it has great physical
transparency and technical simplicity.

Based upon the above reconstruction formulation, 
we can further define and classify the Most Minimal Neutrino Seesaw 
Models (MMNSMs) in which $m_D^{~}$ conatins 2-texture-zeros or
2-equalities or 1-texture-zero plus 1-equality, so that a {\it single}
\CP-phase alone generates all \CP-violations at both high and low energies. 
(Our classification of the MMNSMs include the recent Frampton-Glashow-Yanagida
leptogenesis scheme\,\cite{FGY} as one sub-class.) 
A complete exploration of the MMNSMs will be given elsewhere\,\cite{BDHL2}.

\vspace*{5mm}
\noindent
\underline{{\Large 
$\bullet$}\,{\it Classifying Minimal Schemes for Leptogenesis}}
\vspace*{3mm}

According to the Reconstruction Theorem above, we explicitly classify
our Type-I and Type-II MNSMs. For Type-I class of MNSMs, there are
six 1-texture-zero schemes for \mD in total, 
\beq
\label{eq:Type-I}
\ba{ll}
m_D^{I} = &\!\!
\left\lgroup
\ba{cc}
a &~ 0  \\[1mm]
b &~ b' \\[0mm]
c &~ c'
\ea
\right\rgroup \! ,~~
\left\lgroup
\ba{cc}
a &~ a'  \\[1mm]
b &~ 0   \\[0mm]
c &~ c'
\ea
\right\rgroup \! ,~~
\left\lgroup
\ba{cc}
a &~ a'  \\[1mm]
b &~ b'  \\[0mm]
c &~ 0 
\ea
\right\rgroup \! ,
\\[6mm]
  & \!\!
\left\lgroup
\ba{cc}
0 &~ a'  \\[1mm]
b &~ b'  \\[0mm]
c &~ c'
\ea
\right\rgroup \! ,~~
\left\lgroup
\ba{cc}
a &~ a'  \\[1mm]
0 &~ b'  \\[0mm]
c &~ c'
\ea
\right\rgroup \! ,~~
\left\lgroup
\ba{cc}
a &~ a'  \\[1mm]
b &~ b'  \\[0mm]
0 &~ c' 
\ea
\right\rgroup \!,  
\ea
\eeq
which we label as Type-Ia through Type-If.
For the Type-II MNSMs, there are 15 schemes depending on how the
equality is chosen, which can be horizontal (such as $a=a'$),
or vertical (such as $b=c$), or crossing (such as $b=c'$). 
The Horizontal Equalities (HE) 
in \mD are {\it invariant} under 
rephasing and are thus not affected when we fix the low energy 
parametrization of MNSP matrix in the leptonic charged currents.
The Vertical Equalities (VE) or 
Crossing Equalities (CE) in \mD arising from an 
underlying theory could be changed by the rephasing so they
are less appealing in comparison with the horizontal equalities.
In the following we focus on the analysis of Type-II schemes with
{\it horizontal equality}. There are only 3 such Type-II-HE schemes,
\beq
\label{eq:Type-II}
m_D^{II} = \!\!
\left\lgroup
\ba{cr}
a & \pm a  \\[1mm]
b & b' \\[0mm]
c & c'
\ea
\right\rgroup \!,~~
\left\lgroup
\ba{cr}
a &~~ a'  \\[1mm]
b & \pm b   \\[0mm]
c &~~ c'
\ea
\right\rgroup \!,~~
\left\lgroup
\ba{cr}
a &~~ a'  \\[1mm]
b &~~ b'  \\[0mm]
c & \pm c 
\ea
\right\rgroup \!,
\eeq 
where in each scheme we can have a variation scheme 
by simply flipping the sign of the equality.

\vspace*{5mm}
\noindent 
{\bf 4. Reconstruction Analysis for Leptogenesis 
} 
\vspace*{3mm}

We start by proving a general theorem about the structure 
of the leptogenesis in the two-heavy-neutrino seesaw schemes. 
The theorem will establish the connection between a nonzero
leptogenesis (due to $\ep_1^{~}\neq 0$) and a nonzero solar
mass-squared-difference (due to $\Delta_s\neq 0$).
The {\it Part-I} of this theorem states that 
{\it for any two-heavy-neutrino seesaw with IH, 
nonzero leptogenesis can be realized only if the low energy
observable \,$\D_s$\, is not zero (because  
\,$\ep_1^{~}\to 0$\, as \,$r\equiv \D_s/\D_a \to 0$).}\,
To prove this, we derive, 
for {\it any} two-heavy-neutrino seesaw under \,$r\to 0$\,,
\beq
\label{eq:IH-r0}
\ba{l}
\[\aa^* \aa' + \bb^*\bb' + \cc^*\cc' \]_{\rm IH}
\\[3mm]
= \dis \f{\cxx}{|A|^{2}} \!
\[
\(\aa^*\aa'\!-\!\aa\,\aa'^*\) \!+\!
   \sh_{b'}^{~}\ss\cs \!\(|\aa|^2 \!+\! |\aa'|^2\)\!
   (q^{\f{1}{2}}\!-\!q^{-\f{1}{2}})
\]
\\[5mm]
= \,{\rm Imaginary},\,  ~~\To~~~ \ep_1 = 0\,,
\ea
\eeq
where \,$|A|\neq 0$\, is ensured by the $\nu$-oscillation data.

We next consider the NH scenario. 
Using our reconstruction solution (\ref{eq:RCS1}) and its variations,
and defining 
\,$\Sigma  \dis\equiv [\aa^* \aa' + \bb^*\bb' +\cc^*\cc' ]_{\rm NH}$,\,
we deduce, for \,$r\to 0$\,,
\beq
\label{eq:NH-r0}
\ba{l}
\Sigma \dis
=\f{\,\cc^*\cc'\,}{\,\caa\cxx\,}
=\f{\,\bb^*\bb'\,}{\,\saa\cxx\,}
=\dis\f{\,\aa^*\aa'\,}{\sxx} ,
\ea
\eeq
where the last equality is derived for \,$\sx\neq 0$\,.\, [The combined
limit \,$(r,\,\sx)\to 0$\, will force $(\aa,\,\aa')\to 0$ and 
a vanishing solar angle \,$\TH_s \to 0$\,,\, but this is not our concern
since we examine \,$r\to 0$\, as an independent limit with all mixing
angles confined to their physical ranges.]
Eq.\,(\ref{eq:NH-r0}) shows
that for any $m_D^{~}$ with 1-texture-zero
or 1-horizontal-equaltiy, we have, under the $r\to 0$ limit,
$\,\Sigma=0$\, or  $\,\Sigma= {\rm Real}$,\,
which means $\,\ep_1^{~}=0$\,.\,
Hence, the {\it Part-II} of our theorem states: 
{\it For NH, 
any viable 2-heavy-neutrino seesaw scheme with 1-texture-zero or 
1-horizontal-equaltiy in \mD can have nonzero leptogenesis only if 
the low energy observable $\Delta_s$ is not zero ($r\neq 0$).}

\vspace*{3mm}
\noi
{\bf 4.1. Analysis for Type-I Minimal Schemes
}
\vspace*{2.5mm}

We next turn to explicit analysis of the Type-Ia scheme in 
(\ref{eq:Type-I}) with $a'=0$.   From (\ref{eq:RCS1})
the complete reconstruction solution is given by
\beq
\label{eq:mD-Type-Ia}
\ba{ll}
\dis
\aa =\sh_a\sqrt{A}\,,~~ & \aa' = 0\,,
\\[2mm]
\dis
\bb  = \sh_a\f{C}{\sqrt{A}}\,,~~ & \dis
\bb' = \sh_a \sh_{b'}\sqrt{\f{AB-C^2}{A}} \,,
\\[4mm]
\dis
\cc  =\sh_a\f{D}{\sqrt{A}} \,,~~ & \dis
\cc' =\sh_a \sh_{c'}\sqrt{\f{AF-D^2}{A}} \,.
\ea
\eeq
We thus derive the \CP-asymmetry $\bar{\ep}_1$ 
from (\ref{eq:Epsilon1}),
\beq
\label{eq:Type-1a-ep1}
\bar{\ep}_1 =\dis
\f{\IM\!\[\(C^*\sqrt{AB-C^2}+\sh_{b'}\sh_{c'}D^*\sqrt{AF-D^2}\)^2\]\,}
  {|A|\(|A|^2+|C|^2+|D|^2\)}   ,
\eeq
which is expressed solely in terms of the low energy neutrino observables
[cf. Eqs.\,(\ref{eq:Mnulow}) and (\ref{eq:Mnuij})].

\vspace*{4mm}
\noindent
\underline{{\Large $\bullet$}\,{\it 
Type-I Schemes with Inverted Hierarchy}\,}
\vspace*{3mm}

For the IH scenario, we derive for Type-Ia ($a'=0$),
\beq
\label{eq:IaIH-ep1f}
\epbar = \dis
\f{\sss\css\sqrt{1+r}(1+r\css )^{-1}\,(r\sin\phi )}
  {\[(\css\sqrt{1\!+\!r}+\sss )^2
  +2\sss\css (\cos\phi\! -\! 1)\sqrt{1\!+\!r}\,\]^{1/2}\,} .
\eeq
Strikingly, $\epbar$ depends on Majorana phase
$\phi$ only; it is independent of $\TH_x$ and Dirac phase $\d$.

Type-Id ($a=0$) is very similar to Type-Ia;  
$\epbar$ is given by (\ref{eq:IaIH-ep1f}) 
with the factor $\sqrt{1+r}(1+r\css )^{-1}$ replaced by $-1$.
We have also derived the expressions for other Type-I schemes, which
take simple analytical forms under the expansion in 
$\lambda \equiv O(\sqrt{r},\,\sx )$.
For Type-Ib ($b'=0$) and Type-Ie ($b=0$), we find
\beq
\label{eq:IHb}
\ba{rll}
\epbar &=& \dis
\pm\;\f{\,r\ss\cs\[\ss\cs\ca\sin\phi+2\sa\sx z_1^{~}\]\,}
  {\ca\[1\!+2\sss\css\(\cos\phi\!-\!1\)\]^{1/2}\,}+O(\lambda^4),
\\[6mm]
z_1^{~} & \equiv & \dis
  \css\sin\(\d\!-\!\phi\)\!+\!\sss\sin\(\d\!+\!\phi\),
\ea
\eeq
which are numerically very accurate.
In (\ref{eq:IHb}) the overall $+\,(-)$ sign 
corresponds to Type-Ib\,(-Ie) scheme.

For Type-Ic ($c'=0$) and Type-If ($c=0$), $\epbar$ is given by (\ref{eq:IHb})
with the replacements, $\sa\leftrightarrow \ca$ and $\sx\rightarrow - \sx$.
Interestingly the \CP-asymmetry $\epbar$ is independent of $\TH_x$ 
and the Dirac phase $\d$ (Type-Ia and -Id), 
or depends on them only at higher orders.
This means that in the Type-I schemes with inverted hierarchy, 
leptogenesis is controlled by a single \CP-phase $\phi$  which
also appears in the low energy \nuBB decay observable,
\beq
\label{eq:Mee-IaIH}
\dis
|\MM_{ee}^{~}| = m_0^{~} \cxx \!
\[ 
\( \css\sqrt{1\!+\!r}+\sss\)^2 - 
\sqrt{1\!+\!r}
\sin^2\!2\TH_s\sin^2\!\f{\phi}{2}
\]^{\f{1}{2}} \!\!.
\eeq
Hence the Type-I schemes with IH predict
an {\it unique link} between high energy
and low energy \CP-violations via a {\it single phase $\phi$.}
But, given the anticipated future uncertainty in the nuclear matrix 
elements\,\cite{0nuBB,0nuBB2}, the small size of the predicted $|\MM_{ee}|$, and 
the expected uncertainty in the eventual cosmological determination of 
the sum of neutrino masses for the hierarchical neutrino spectrum\,\cite{Sum}, 
it is hard to probe the Majorana phase $\phi$ \cite{barglash2}.
Our study of the low energy test of leptogenesis
provides a motivation for greatly improving
the $|\MM_{ee}|$ determination;
such efforts are currently underway\,\cite{0nuBB2}.

\vspace*{4mm}
\noindent
\underline{{\Large $\bullet$}\,{\it 
Type-I Schemes with Normal Hierarchy}\,}
\vspace*{3mm}

For NH scenario, we derive for Type-Ia,

\beq
\label{eq:IaNH-ep1f}
\ba{rll}
 & & \\[-10mm]
\epbar &=& \dis
\f{\,\sss\sxx\cxx\sqrt{r(1+r)}\,\sin(\phi+2\d )\,}
  {\,\[r(1-\css\cxx )+\sxx \]z_2^{~}\,} 
\\[5mm]
&=& \dis
\f{\sss\sxx\sin(\phi+2\d )}
  {\,\[r\sss +\sxx\]\[\sss +\cos(\phi+2\d )
      \sxx /\sqrt{r}\,\]\,}+O(\lambda^3) , 
\\[6mm]
z_2^{~} & \equiv & \dis \[
 \( \cxx\sss \sqrt{r}+\sxx\sqrt{1+r}\,\)^2 + \right.
\\[2mm]
&& 
\left. \dis
  2\sqrt{r(1+r)} \sss\sxx\cxx
\(\cos(\phi+2\d )-1\)\]^{\f{1}{2}} ,
\ea
\eeq
where $\lambda \equiv O(\sqrt{r},\,\sx )$.
Type-Id is again very similar to Type-Ia except that 
in (\ref{eq:IaNH-ep1f}) the denominator of $\epbar$ will have
its first brackets $[\cdots ]$ replaced by $-[\sss \!+\!\css\sxx ]$.
Unlike the IH case, the $\epbar$
is now controlled by a single phase-combination
$\phi+2\d$ from low energy. Interestingly, we find that the
\nuBB decay observable now becomes
\beq
\label{eq:Mee-IaNH}
\dis
|\MM_{ee}^{~}| = m_0^{~}  \!
\[ 
\sss\sqrt{r}+\sxx\cos\!\(\phi+2\d\) + O(\lambda^3)
\] \,.
\eeq
which depends on the {\it same} phase-combination $\phi+2\d$.
Furthermore, the Dirac phase $\d$  can be measured at low energy 
in long-baseline neutrino oscillation experiments 
via the leptonic Jarlskog invariant\,\cite{jarlskog},
%
\beq
\label{eq:J}
{\cal J}=\f{1}{8}\sin\!2\TH_s\sin\!2\TH_a\sin\!2\TH_x\cos\!\TH_x
\sin\!\d \,.
\eeq

The results of $\bar\epsilon_1$ in
other Type-I schemes are not so simple for the normal hierarchy. 
For Type-Ib, we find
\beq
\label{eq:NHb}
\epbar = \dis
\f{\sqrt{r}\cs\ca\[\cs\ca\sin\phi\!+\!2\ss\sa\sx\sin\(\phi\!+\!\d\)\]}
  {\saa\!+\!\sqrt{r}\css\caa\cos\phi}+O(\lambda^3),
\eeq
while for Type-Ie, $\epbar$ is given by
\beq
\label{eq:NHe}
\f{-\saa\cs\ca\[\cs\ca\sin\phi\!+\!2\ss\sa\sx\sin\(\phi\!+\!\d\)\]}
  {\[\css\!+\!\sss\saa\!+\!2\ss\cs\sa\ca\sx\cos\!\d\]
  \!\[\saa +\!\sqrt{r}\css\caa\cos\!\phi\]}
  +O(\lambda^2).
\eeq
For $c'=0$ (Type-Ic) or $c=0$ (Type-If), $\epbar$ is given by (\ref{eq:NHb})
or (\ref{eq:NHe}) with the replacements
\,$\sa\leftrightarrow\ca$\, and  \,$\sx\to -\sx$.\,
%

We may mention that the phase convention below (\ref{eq:Mnuij}) 
is not unique. If we assign the independent Majorana phase instead to the 
mass $m_3'$ (NH), or $m_1'$ (IH), the results for $\epbar$ would change only 
by $\phi \rightarrow -\,\phi$ everywhere.

Finally we analyze the predictions for leptogenesis 
in the Type-Ia schemes [cf. Fig.\,1].
For the IH scenario, we first show the asymmetry $\eta_B^{~}$ versus
the low energy Majorana \CP-phase $\phi$, where
we set  $x=M_2/M_1=10$.   [The result is very insensitive 
to $x$ for \,$x\geqq 3$,\,  because in the SM and MSSM,
\,$xF(x)$\, equals $\,-\f{3}{2}~{\rm and}~-3$,\,  
to 7\%~(2\%) accuracy or better for $x\geqq 3~(x\geqq 5)$.]
Fig.\,1(a) shows that for the best fit values of $\nu$-oscillation
parameters the CMB limit (1) 
constrains $M_1 \gtrsim 10^{13}$\,GeV for IH,  and for 
\,$M_1 \geqq 1.3\times 10^{13}$\,GeV  the phase $\phi$ is confined 
to two narrow regions.
In Fig.\,1(b),  we plot the physical mass $M_1$ vs.\ the low energy observable 
$|\MM_{ee}|$ of \nuBB decay which
are both controlled by the same \CP-phase $\phi$\,.\, 
Since $\ep_1$ is sensitive to $(r,\,m_0^{~})$, 
we have scanned the 95\%\,C.L. range 
of \,$(\Delta_s,\,\Delta_a)$ (and also $\TH_s$)  
in addition to varying $\phi$, and found that $M_1$ is always 
above \,$4\times 10^{12}$\,GeV,\, but has little chance to go  
above  \,$2\times 10^{14}$\,GeV.  Future improved precision of
$\,(\Delta_s,\,\Delta_a)$\,  will strengthen the bound.

For the NH scenario, we first plot $\eta_B^{~}$ as a function of
$\sx$ by varying the \CP-phase angle $\phi+2\d$ [cf. Fig.\,1(c)].  
Then, in Fig.\,1(d) we analyze the correlation between the scale $M_1$ 
and the low energy CP-observable $J$ in (\ref{eq:J}) by varying
both phases $(\phi,\,\d)\in (0,\,2\pi)$. 
We have scanned the 95\%\,C.L. ranges of all $\nu$-oscillation parameters, 
together with $\eta_B^{~}$.\,
Distinctly, we find  \,$M_1\geqq 3\times 10^{10}$\,GeV, 
and for \,$|J|\geqq 0.01$,\, 
$M_1$ is mainly below  $10^{12}$\,GeV.

\begin{figure*}
\centerline{
\includegraphics[width=9.6cm,height=6.0cm]{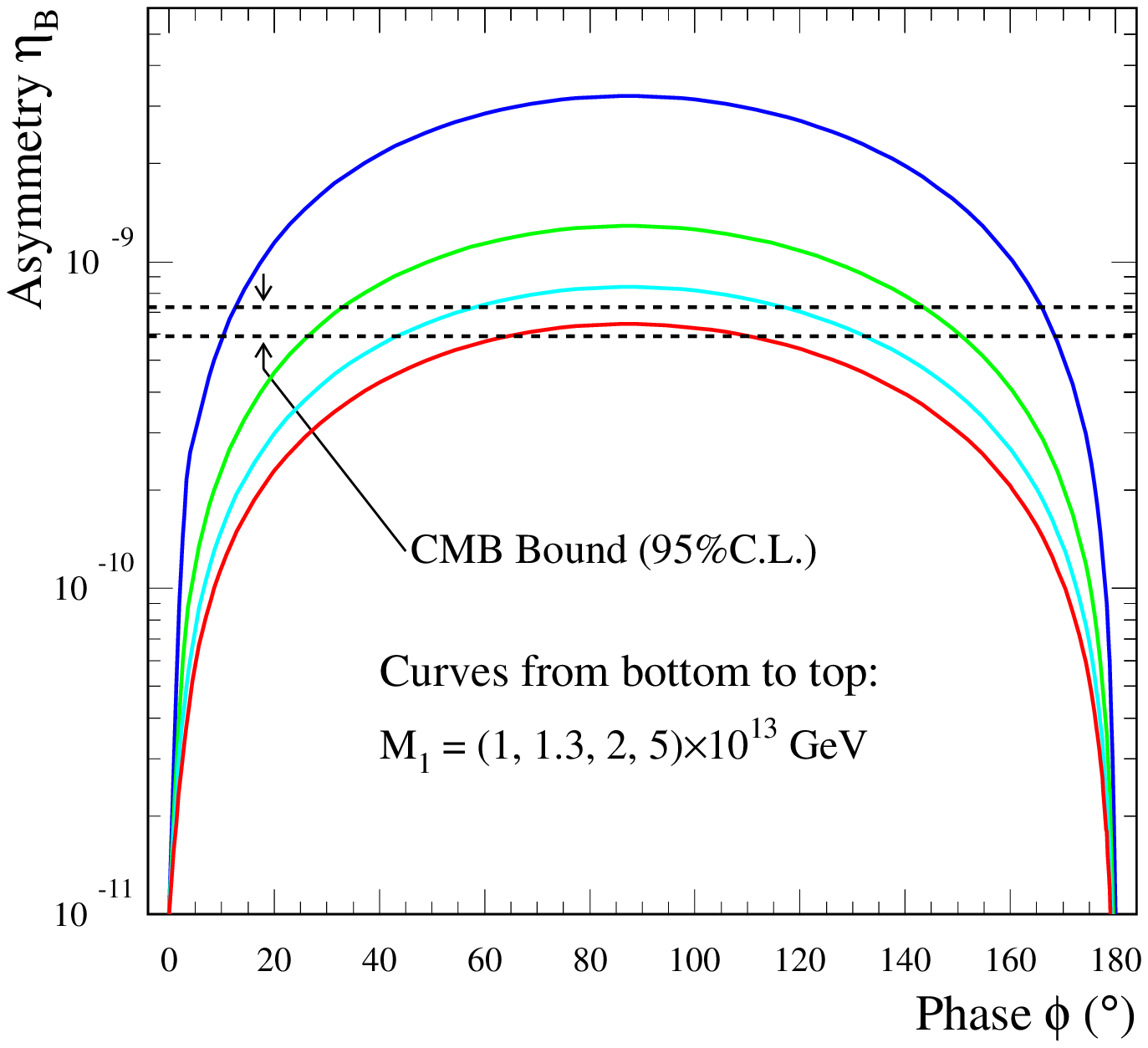}  
\hspace*{-7mm}
\includegraphics[width=8.8cm,height=6.0cm]{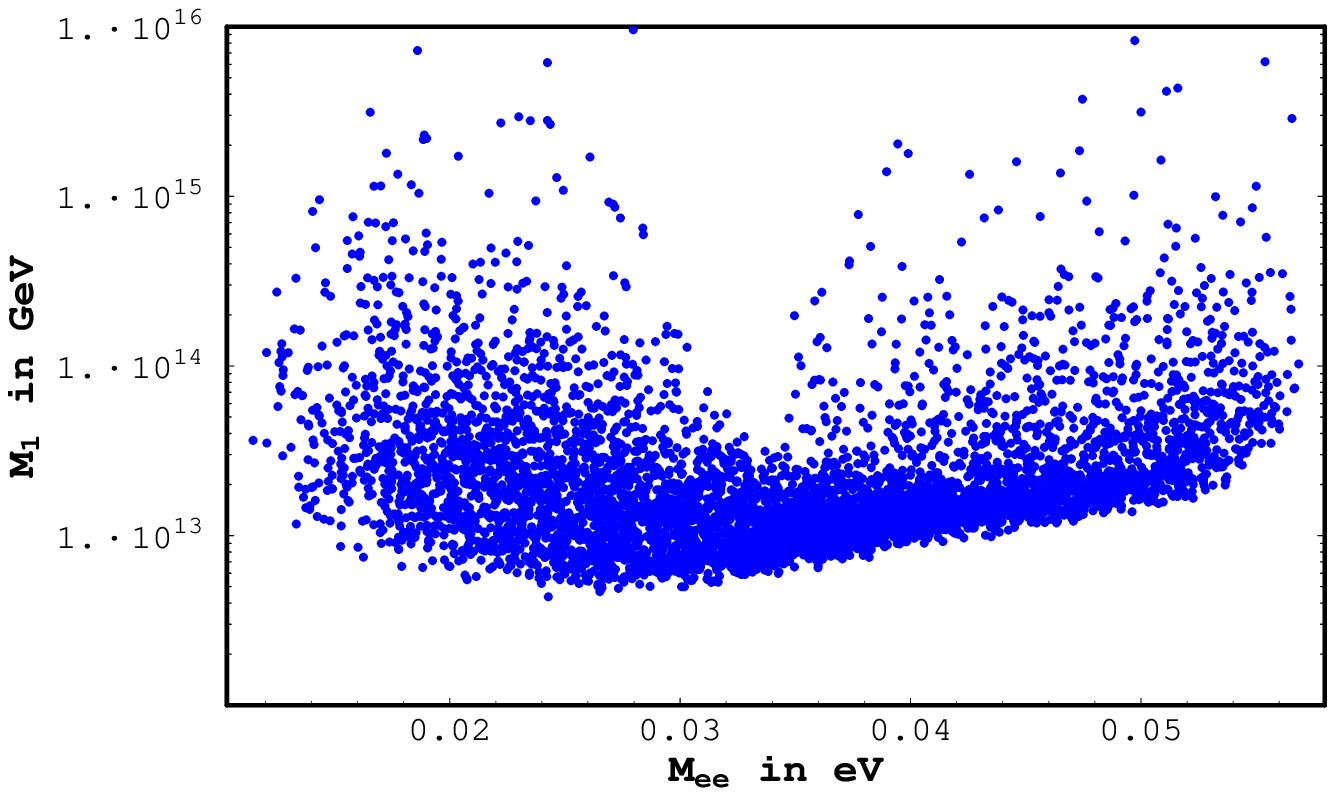}
}
\centerline{
\includegraphics[width=9.6cm,height=6.0cm]{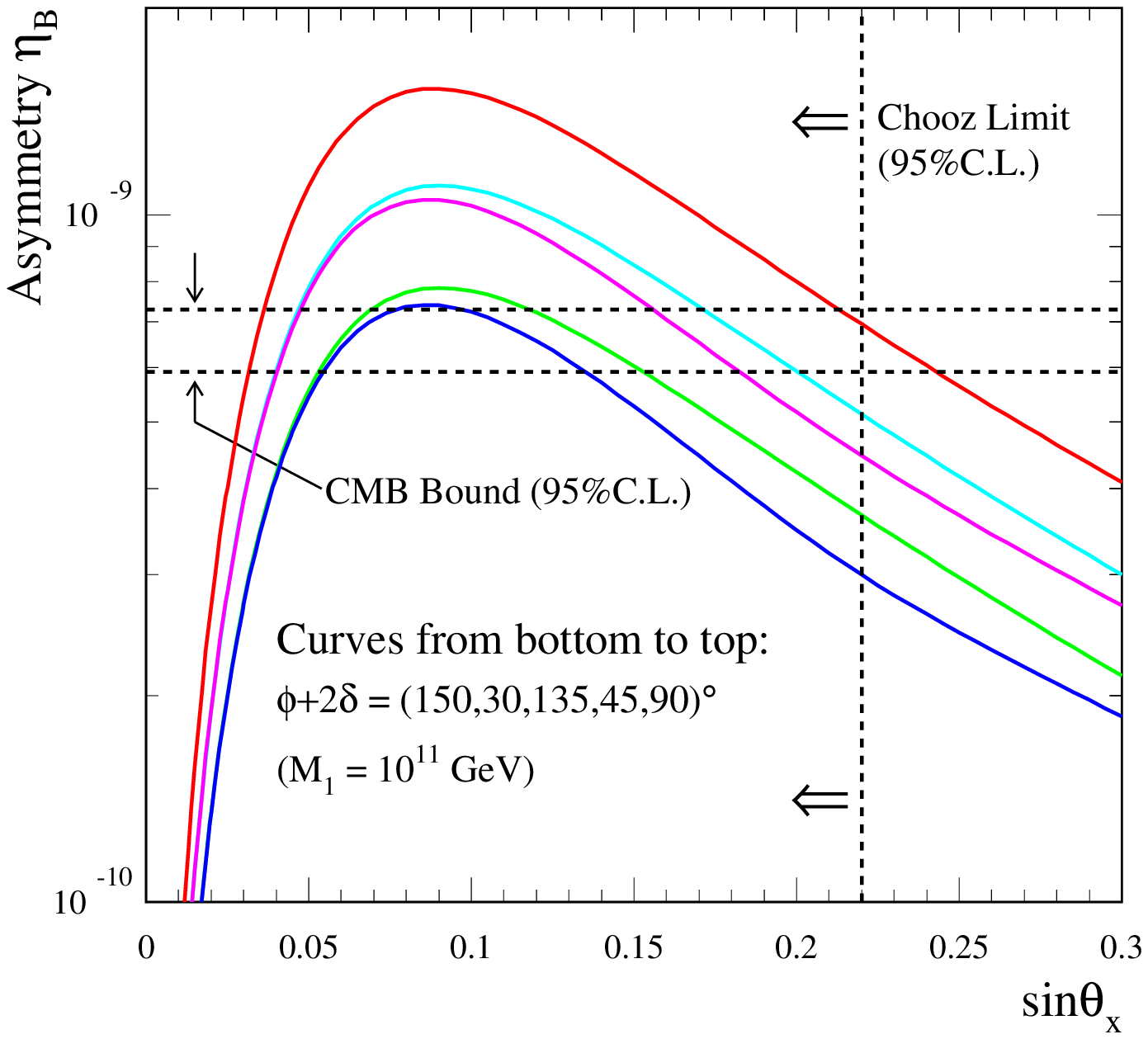}  
\hspace*{-7mm}
\includegraphics[width=8.8cm,height=6.0cm]{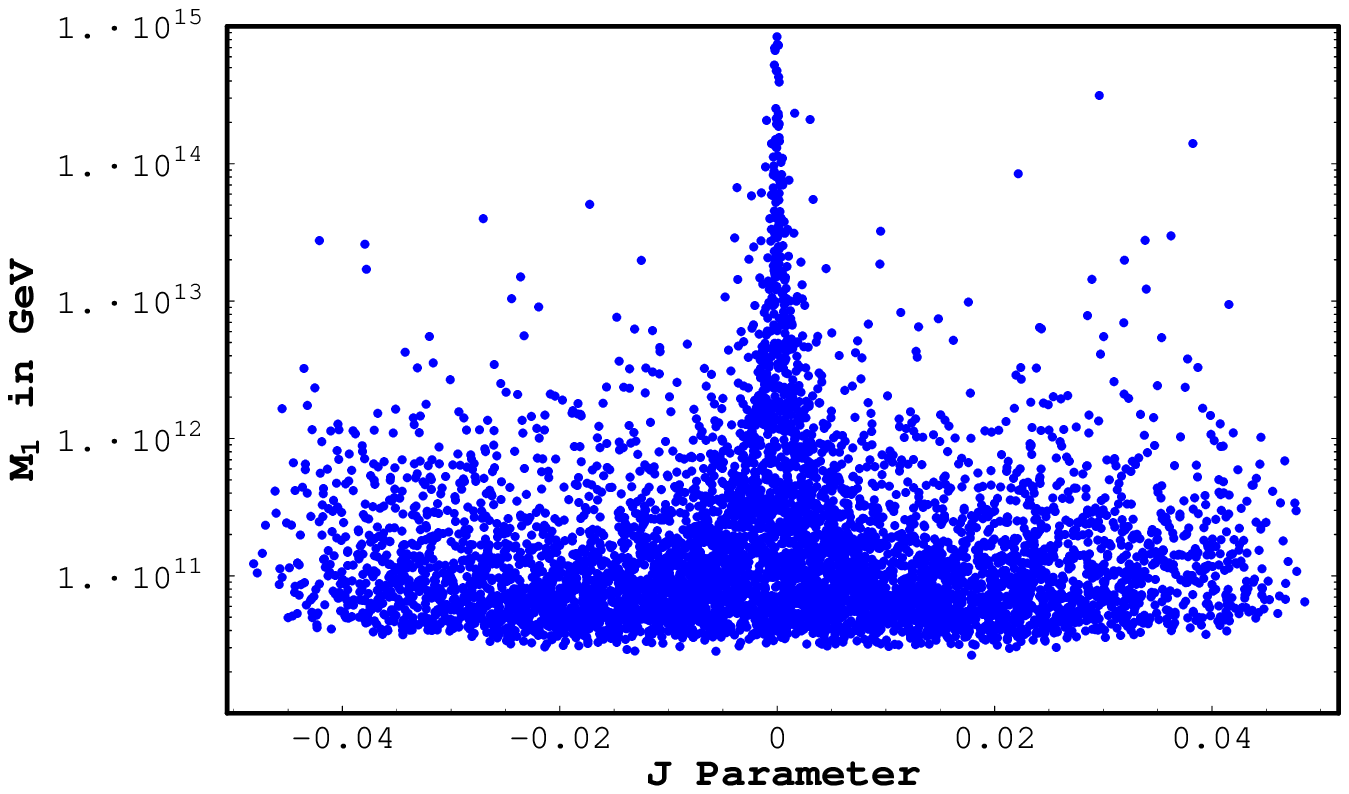}
}
\setlength{\unitlength}{1mm}
\begin{picture}(0,0)
\put(-76,115){(a)}
\put(18,115){(b)}
\put(-76,56){(c)}
\put(18,56){(d)}
\end{picture}
\vspace*{-3.5mm}
\caption{Analysis for Type-Ia.
(a)~Baryon asymmetry $\eta_B^{~}$ vs.~the 
    low energy Majorana phase $\phi$ in the IH scenario,
    with best-fit values of $\nu$-oscillation observables as inputs. 
(b)~Leptogenesis scale $M_1$ vs.~the $0\nu\beta\beta$-decay 
    observable $M_{ee}\,(\equiv |\MM_{ee}|)$ with IH, where 
    we have varied the phase $\phi\in (0,\,2\pi )$,  
    imposed the 95\%\,C.L. CMB limit and scanned the
    95\%\,C.L. ranges of $(\Delta_s,\,\Delta_a,\,\TH_s,$ $\,\sx)$, for 
    $1.2\times 10^4$ samples.
    (Here $\sx$ is relevant for $|\MM_{ee}|$, but varying $\sx$
    within $\,|\sx | \leqq 0.22$ has only minor effect on  $|\MM_{ee}|$
    because $\cxx > 0.95\simeq 1$.)
(c)~Asymmetry $\eta_B^{~}$ vs.~$\sin\theta_x^{~}$ 
    in the NH scenario,
    with best-fit values of the $\nu$-oscillation observables as inputs. 
(d)~Mass scale $M_1$ vs.~the \CP-violation observable $J$ in the NH scenario, 
    where we have varied $(\delta,\,\phi)\in (0,\,2\pi)$,  
    imposed the CMB limit and scanned the $\nu$-oscillation parameters
    in their 95\%\,C.L. ranges, for $1.2\times 10^4$ samples.
}
\label{fig:1}
\vspace*{-3mm}
\end{figure*}

\vspace*{4mm}
\noindent 
{\bf 4.2. Analysis for Type-II Minimal Schemes
}
\vspace*{3mm}

Next we turn to the analysis of the Type-II schemes with horizontal
equality. We consider the Type-II-HEa Schemes with the
equality  $a=a'$.
From (\ref{eq:RCS1})
we thus derive a complete reconstruction solution,
\beq
\label{eq:mD-Type-IIHEa}
\ba{l}
\dis \aa  = \sh_a\f{\sqrt{A}}{~\omega_2}\,,~~~~~~
\dis \aa' = \sh_a\f{\sqrt{A}}{~\omega_1}\,,
\\[5mm]
\dis\bb =\sh_a\f{1}{\,\omega_2\sqrt{A}\,}
\[C-\sh_{b'}  \rho^{-1}\sqrt{AB-C^2}\]\! ,
\\[5mm]
\dis\bb' =\sh_a\f{1}{\,\omega_1\sqrt{A}\,}
\[C + \sh_{b'}\rho \,\sqrt{AB-C^2}\]\! ,
\\[5mm]
\dis\cc =\sh_a\f{1}{\,\omega_2\sqrt{A}\,}
\[D- \sh_{c'} \rho^{-1}\sqrt{AF - D^2}\]\! ,
\\[5mm]
\dis\cc' =\sh_a\f{1}{\,\omega_1\sqrt{A}\,}
\[D+ \sh_{c'} \rho \,\sqrt{AF - D^2}\]\! ,
\ea
\eeq
where we have introduced the notation 
$\rho \equiv \sqrt{M_2/M_1}$,
$\omega_1=\sqrt{1+\rho^2}$, and $\omega_2 =\sqrt{1+\rho^{-2}}$.  
The \CP-asymmetry observable $\epbar$ is given by
\beq
\label{eq:Type-IIHEa-ep1}
\epbar = \dis
\f{\,2(\rho +\rho^{-1})\,}{\omega_1^2|A|}
\f{\sh_{b'}X+(\rho - \rho^{-1})Y_1}
  {X_1+\rho^{-2}X_2-\sh_{b'}2\rho^{-1}Y_1}\,Y_2\,,
\eeq
where 
\beq
\ba{lcl}
X & \equiv & X_1-X_2
\\[2mm]
  & =& \dis
\[|A|^2+|C|^2+|D|^2\]
-[|AB-C^2|+|AF-D^2| ],
\\[3mm]
Y &\equiv& Y_1+i\,Y_2\,,
\\[2mm]
 & = & \dis
C^*\sqrt{AB-C^2} \,+\, \sh_{b'}\sh_{c'}D^*\sqrt{AF-D^2}\,,
\ea
\eeq
and
\,$\sh_{b'}\sh_{c'}=+\,(-)$\, for IH~(NH) Schemes.

\vspace*{4mm}
\noindent
\underline{{\Large $\bullet$}\,{\it 
Type-II Schemes with Inverted Hierarchy}\,}
\vspace*{3mm}

For the IH scenario, we derive the \CP-asymmetry of leptogenesis
for Type-II-HEa Schemes,
\beq
\label{eq:IIaIH-ep1f}
\epbar = \dis
\f{2\ss\cs\!\[\sh_{b'}(\css\!-\! \sss )
            +(\rho\!-\!\rho^{-1})\ss\cs\cos\f{\phi}{2}\]}
  {(\rho + \rho^{-1})
   \[ 1-4\sss\css\sin^2\!\f{\phi}{2}\]^{1/2}}\(r\sin\!\f{\phi}{2}\)\!,
\eeq
where we have expanded the exact formula up to $O(r)$ and ignored small
sub-leading terms of $O(r^2)$.\,
Again we see that for Type-II-HEa Schemes with IH, the \CP-asymmetry
$\epbar$ only contains the Majorana phase $\phi$; it is
independent of $\sx$ and the Dirac phase $\d$.\,
A new feature is the nontrivial dependence on 
\,$\rho=\sqrt{M_2/M_1}$\,
for \,$\rho\neq 1$.\,

\begin{figure*}
\centerline{
\includegraphics[width=9.6cm,height=6.0cm]{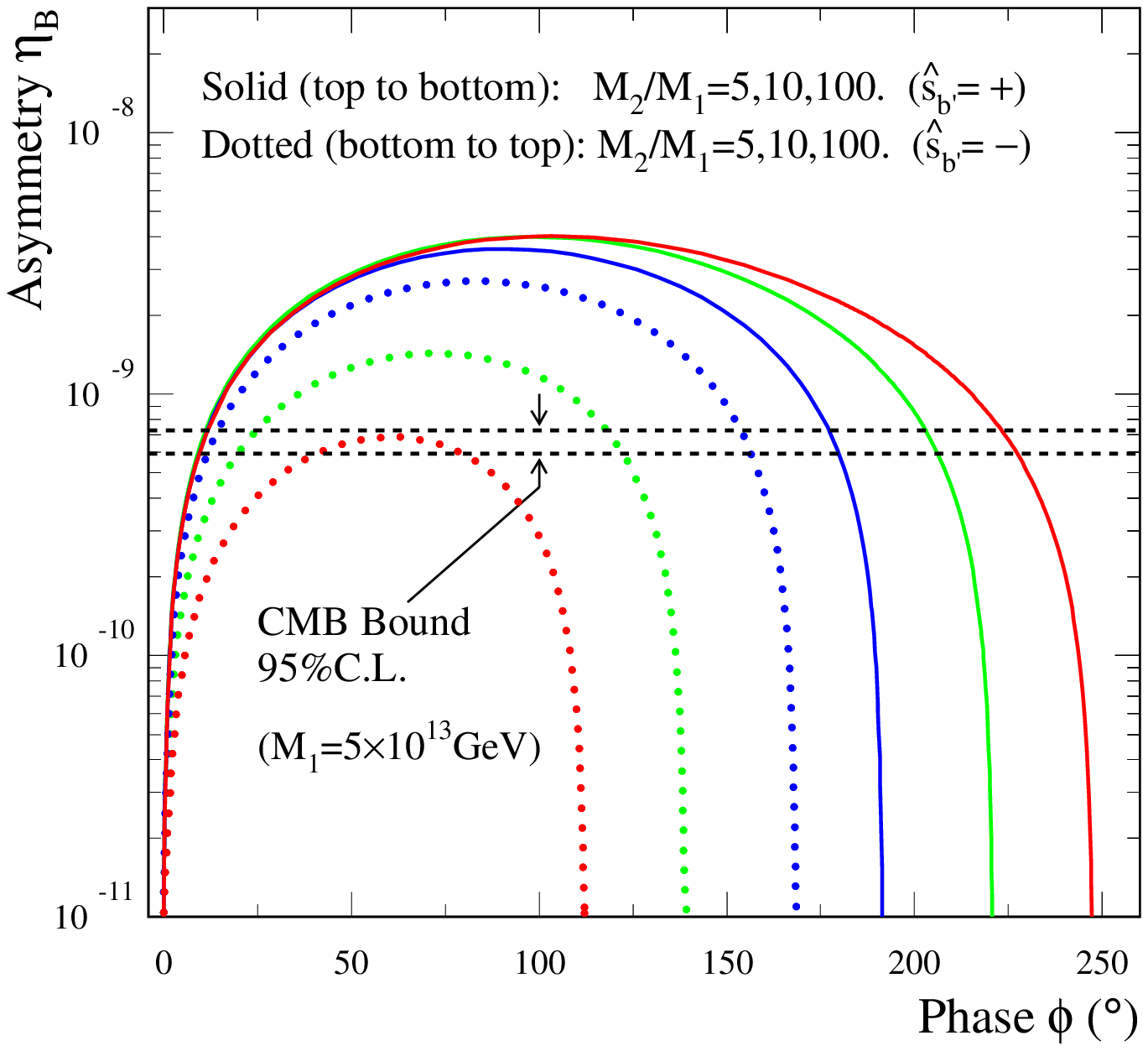}  
\hspace*{-7mm}
\includegraphics[width=8.8cm,height=6.0cm]{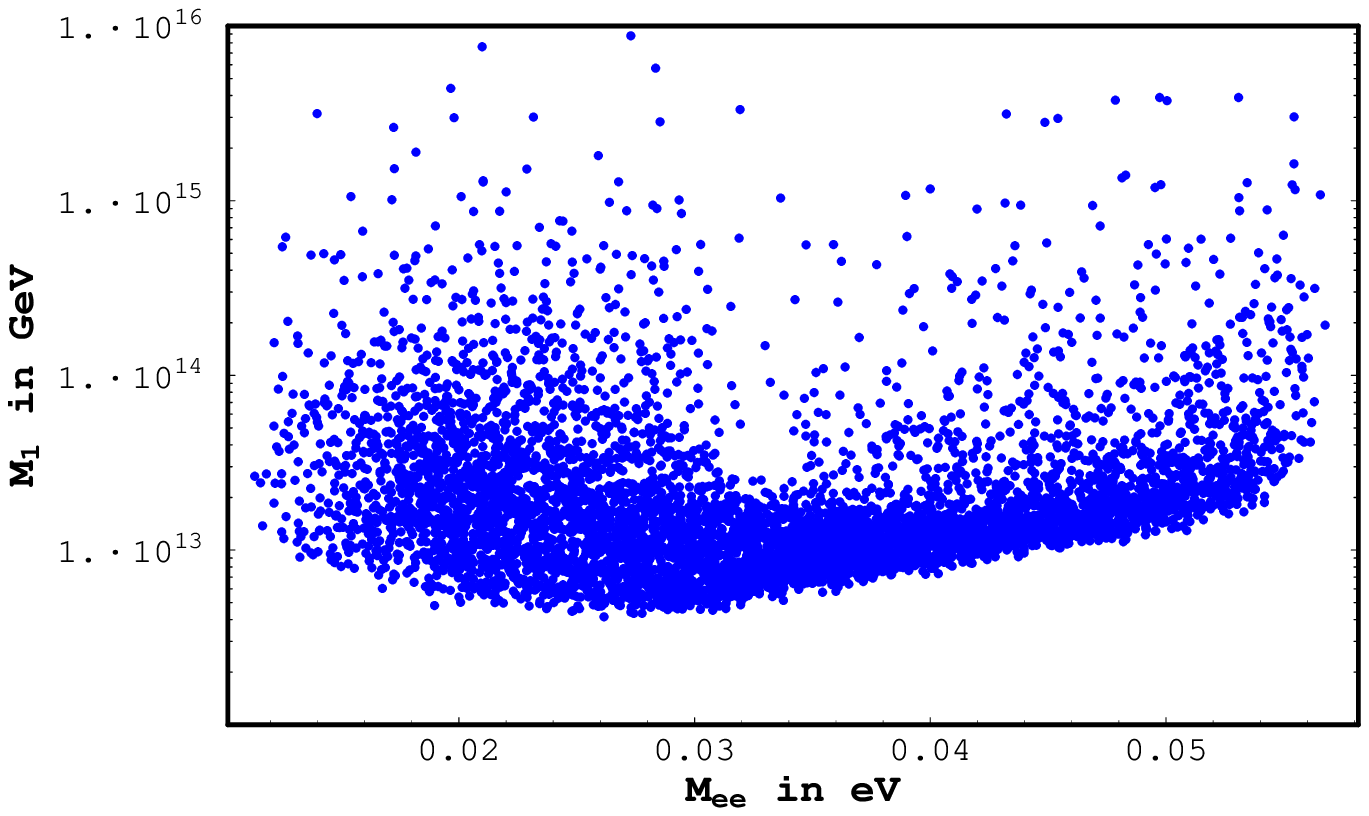}
}
\centerline{
\includegraphics[width=9.6cm,height=6.0cm]{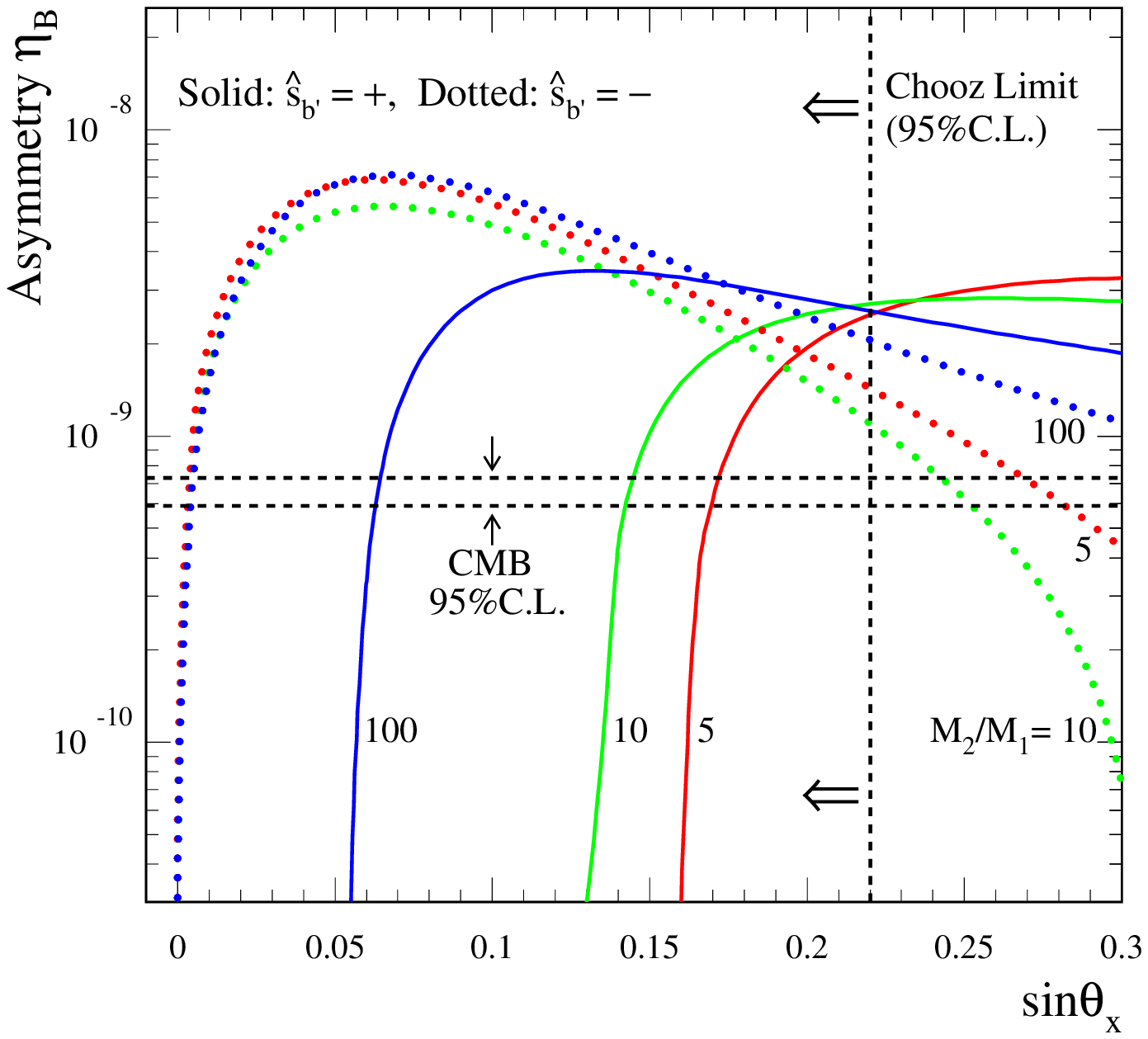}  
\hspace*{-7mm}
\includegraphics[width=8.8cm,height=6.0cm]{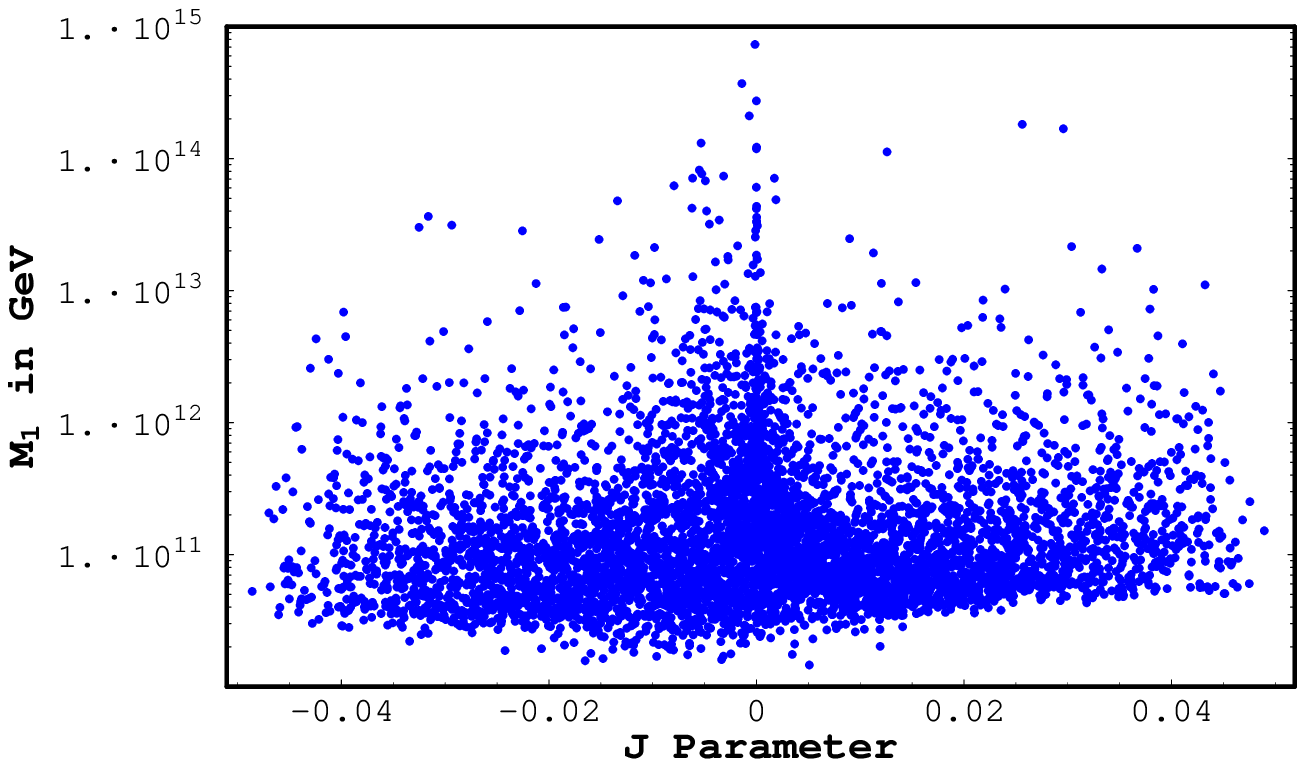}
}
\setlength{\unitlength}{1mm}
\begin{picture}(0,0)
\put(-3,76){(a)}
\put(10.2,76){(b)}
\put(-3,55){(c)}
\put(10.2,55){(d)}
\end{picture}
\vspace*{-4mm}
\caption{Analysis for Type-II-HEa.
(a)~Baryon asymmetry $\eta_B^{~}$ vs.~the low 
     energy Majorana phase $\phi$ in the IH scenario,
     with best-fit values of $\nu$-oscillation observables as inputs; 
     different curves show the significant effects of varying the ratio
     \,$M_2/M_1$\, and the sign $\sh_{b'}$.
(b)~Leptogenesis scale $M_1$ vs.~the $0\nu\beta\beta$-decay 
     observable $M_{ee}\,(\equiv |\MM_{ee}|)$ with IH
     (for $\sh_{b'}=+$),  where 
     we have varied $\phi\in (0,\,2\pi )$ and $M_2/M_1\in (5,\,100)$,
     imposed the 95\%\,C.L. CMB limit and scanned
     the 95\%\,C.L. ranges of $(\D_s,\,\D_a,\,\TH_s,\,\sx )$, 
     with  $1.2\times 10^4$  samples.
(c)~Asymmetry $\eta_B^{~}$ vs.~$\sin\theta_x^{~}$ 
     in the NH scenario,
     with best-fit values of the $\nu$-oscillation observables as inputs;
     also \,$M_1=5\times 10^{11}$\,GeV  
     and \,$\phi+2\delta = 135^\deg$\, are chosen
     for illustration.
(d)~Mass scale $M_1$ vs.~\CP-violation observable $J$ in the NH scenario 
     (for $\sh_{b'}=+$), 
     where we have varied $(\delta,\,\phi)\in (0,\,2\pi)$,  
     and scanned the 95\%\,C.L. ranges of the $\eta_B^{~}$ and
     $\nu$-oscillation parameters, with 
     $1.2\times 10^4$ samples.
}
\label{fig:2}    
\vspace*{-3mm}
\end{figure*}

\vspace*{4mm}
\noindent
\underline{{\Large $\bullet$}\,{\it 
Type-II Schemes with Normal Hierarchy}\,}
\vspace*{3mm}

For the NH scenario, we derive the \CP-asymmetry of leptogenesis
for Type-II-HEa Schemes,
\beq
\label{eq:IIaNH-ep1f}
\ba{lcl}
\epbar 
& =& \dis\f{~ -\sh_{b'}\ss r^{1/4}
     \,+\, (\rho-\rho^{-1})\sx\cos\f{\phi+2\d}{2} ~}
{~\rho (r\sss+\sxx)+\rho^{-1}\sss \sqrt{r}
 -\sh_{b'}2r^{1/4}\ss\sx\!\cos\f{\phi+2\d}{2}\,}
\\[5mm]
& & \dis
       ~\times {\,2\,\sx\sin\f{\phi+2\d}{2}~} 
       +O(\lambda^2)  \,.
\ea
\eeq
A similar feature to Type-Ia with NH [cf. (\ref{eq:IaNH-ep1f})] 
is that the \CP-asymmetry $\epbar$ in 
(\ref{eq:IIaNH-ep1f}) also solely depends
on the phase-combination  $\phi+2\d$.

We have analyzed the prediction of baryon asymmetry $\eta_B$ 
in Type-II-HEa schemes.
For the IH scenario, we first plot the asymmetry $\eta_B$ versus
the low energy Majorana phase $\phi$ in Fig.\,2(a).
Unlike the Type-I schemes 
[cf. Eqs.\,(\ref{eq:Type-1a-ep1})-(\ref{eq:IaIH-ep1f})], 
we find $\eta_B^{~}$ to be very sensitive to the mass ratio
$\rho^2=M_2/M_1$ and the sign $\sh_{b'}$,  according to 
Eqs.\,(\ref{eq:Type-IIHEa-ep1}) and (\ref{eq:IIaIH-ep1f}).
We show, in Fig.\,2(b), the correlation between the leptogenesis 
scale $M_1$ and the \nuBB decay observable $|\MM_{ee}|$ where
we have varied the mass ratio \,$M_2/M_1=5-100$\, in addition to scanning 
all $\nu$-oscillation parameters within 95\% ranges. We find that 
$M_1$ is always above $4\times 10^{12}$\,GeV and largely
falls below $10^{14}$\,GeV.

For the NH scenario, Fig.\,2(c) plots $\eta_B^{~}$ versus 
$\sin\theta_x^{~}$ with different inputs of the ratio $M_2/M_1$ and
sign $\sh_{b'}$, which can be compared with the Type-Ia in Fig.\,1(c).
Fig.\,2(d) demonstrates the correlation between
the mass $M_1$ and the $J$ parameter by scanning $M_2/M_1\in (5,\,100)$
and all other observables within 95\%\,C.L. ranges.
Unlike Fig.\,1(d), the distribution of $M_1$ has no clear peak around \,$J=0$\,.

Finally, we comment on the supersymmetry prediction for $\eta_B^{~}$.
SUSY approximately doubles $g_{*1}$ and the function
$F(x)$  so that their ratio remains
about the same.  The conversion factor 
\,$\xi/(\xi-1)\simeq -0.53$\, 
[cf. (\ref{eq:YB})] is very close to the SM value $-0.55$. 
Another effect is the factor  \,$\sin^2\!\beta < 1$\,
which arises from the 2-Higgs-doublet structure of 
the MSSM but is nearly 1 for $\tan\!\beta\geqq 5$.
We also note that for light neutrinos with a
hierarchical mass-spectrum, possible renormalization
group running effects for the neutrino mixing angles and phases are 
generally small\,\cite{RGnu}, and can be directly incorporated in our
reconstruction analysis.

\vspace*{5mm}
\noindent    
{\bf 5. Summary and Extension}
\vspace*{3mm}

In this Letter 
we have presented a general formalism to 
quantitatively reconstruct leptogenesis for the observed 
matter-antimatter asymmetry (1) in the minimal neutrino seesaw.
We have systematically classified and analyzed 
such minimal seesaw schemes in which {\it the
required high energy \CP-asymmetry for leptogenesis is  uniquely
linked to the low energy \CP-phases in the physical MNSP matrix.}
Imposing the cosmological bound (1) and the existing neutrino oscillation
data, we have analyzed the constraints and correlations of the 
leptogenesis scale $M_1$
versus the low enery neutrino observables $|\MM_{ee}|$ (for \nuBB decays)
and $J$-invariant (for long baseline $\nu$-oscillations).
The minimal schemes of Type-I (with one-texture-zero) and 
Type-II (with one-equality) result in distinctive predictions for
leptogenesis and its link to the low energy \CP-violation
observables (cf. Figs.\,1-2).

We can readily extend our reconstruction formalism to
the general three-heavy-neutrino seesaw where
\,${\cal N}=(N_1,\,N_2,\,N_3)^T$\, and 
\,$M_R={\rm diag}(M_1,\,M_2,\,M_3)$\, 
in their mass-eigenbasis. 
Consequently the Dirac mass matrix $m_D^{~}$ becomes
a $3\times3$ matrix,

\beq
\label{eq:mD3x3}
\ba{l}
\\[-10.5mm]
m_D^{~} ~= \dis 
\left\lgroup 
\ba{ccc} 
a~ & a' & a'' \\[2mm]
b~ & b' & b'' \\[1mm]
c~ & c' & c''
\ea
\right\rgroup
=  
\left\lgroup
\ba{ccc}
\zeta_1\aa~ & \zeta_2\aa'~ & \zeta_3\aa'' \\[1.4mm]
\zeta_1\bb~ & \zeta_2\bb'~ & \zeta_3\bb'' \\[1mm]
\zeta_1\cc~ & \zeta_2\cc'~ & \zeta_3\cc''  
\ea
\right\rgroup \! ,
\ea
\eeq
where \,$\zeta_j \equiv \sqrt{m_0M_j}$\, ($j=1,2,3)$,\,
and  $m_D^{~}$ contains 3 extra independent \CP-phases in
$(\aa'',\,\bb'',\,\cc'')$.  
A general classification of minimal schemes for  
(\ref{eq:mD3x3}) and a complete reconstruction of leptogensis
will be given elsewhere. 
Here we comment on the thermal leptogenesis with
a hierarchical mass-spectrum of ${\cal N}$ such as,
%
$\,
M_1 \ll M_2 \ll M_3 
$,\,
%
which is sometimes 
called light sequential dominance\,\cite{king2}.
With this mass hierarchy, leptogenesis arises 
from $\ep_1$ via the $N_1$ decays, which is extended from
(\ref{eq:Epsilon1}) as
\beq
\label{eq:Epsilon1N3}
\ba{l}
\ep_1^{~} = \dis\f{m_0^{~}M_1}{8\pi}\(\f{\sqrt{2}}{v\sin\!\beta}\)^2\!
             \[ 
x_{21}^{~}F\!\(x_{21}^{~}\)\ov{\ep}_{12} +
x_{31}^{~}F\!\(x_{31}^{~}\)\ov{\ep}_{13}
             \],
\\[5mm]
\ov{\ep}_{13}^{~}  \equiv\!  \dis\f{\Im\m\[\((m_D^\dag m_D^{~})_{13}\)^2\]}
                            {(m_0^{~}M_3)(m_D^\dag m_D^{~})_{11}} 
=\!\dis\f{\Im\m\[(\aa^*\aa''\!+\!\bb^*\bb''\!+\!\cc^*\cc'')^2\]}
            {|\aa|^2+|\bb|^2+|\cc|^2} ,
\ea
\eeq
%
where the formula for
$\,\ov{\ep}_{12}^{~}\,$ takes the same form as $\,\epbar$\, 
in (\ref{eq:Epsilon1}), 
and $\,(x_{21}^{~},\,x_{31}^{~} ) \equiv (M_2/M_1,\,M_3/M_1)$.
The entries in the Dirac mass matrix \mD are generally 
of $O(v)$, except for the possible texture zeros. 
Thus the mass hierarchy \,$M_1 \ll M_2 \ll M_3$\,
implies
\beq
\ba{l}
\dis
\f{\,\,\ov{\ep}_{13}^{~}\,}{\,\,\ov{\ep}_{12}^{~}} 
= \f{~O(\aa'',\,\bb'',\,\cc'')^2~}{O(\aa',\,\bb',\,\cc')^2} 
= O\!\(\!{\f{M_2}{M_3}}\) \ll 1 \,.
\ea
\eeq
Hence, with the above mass hierarchy,
the \CP-asymmetry \,$\ep_1$\, is dominated
by \,$\ov{\ep}_{12}$\, (contained in the 2-neutrino seesaw),
with the \,$\ov{\ep}_{13}$\, term suppressed by the factor  
\,$M_2/M_3\ll 1$.\,
Furthermore, the modifications to the reconstruction solutions
of $(\aa,\,\bb,\,\cc)$ and $(\aa',\,\bb',\,\cc')$ relative to 
a minimal $(N_1,N_2)$-seesaw are also suppressed by $M_2/M_3\ll 1$.
Thus with \,$M_1 \ll M_2 \ll M_3$,\,
leptogenesis effectively reduces to that of
the minimal seesaw in Sec.\,2-4,
\beq
\label{eq:Ep1expand}
\ep_1 = \dis\f{m_0^{~}M_2}{8\pi}\(\f{\sqrt{2}}{v\sin\!\beta}\)^{\!2}\!
F\!\(\f{M_2}{M_1}\)
\ov{\ep}_{12}\[1 + O\!\(\f{M_2}{M_3}\)\] \! , 
\eeq   
which allows the direct application of our Minimal Seesaw Schemes 
(Sec.\,2-4) to the 3-neutrino seesaw at the leading order
of the \,$M_2/M_3$\, expansion.

\vspace*{4mm}
\noindent
{\bf Acknowledgments.}~\,This work was supported by the DOE  
under grants DE-FG02-95ER40896 and DE-FG03-93ER40757,
by the NSF under grant PHY-0070928, and by the Wisconsin Alumni
Research Foundation.


\end{document}